\documentstyle[12pt,epsf]{article}

\newcommand{\RR}{\mbox{${\rm \:  R\!\!\!\! I
\;\;}$}}

\newcommand{\vs}{\vspace{0.25cm}}

\newtheorem{theorem}{Theorem}
\newtheorem{itlemma}{Lemma}[section]
\newtheorem{itproposition}[itlemma]{Proposition}
\newtheorem{itcorollary}[itlemma]{Corollary}
\newtheorem{itremark}[itlemma]{Remark}
\newtheorem{itremarks}[itlemma]{Remarks}
\newtheorem{itdefinition}[itlemma]{Definition}
\newtheorem{itexample}[itlemma]{Example}

\newenvironment{lemma}{\begin{itlemma}\rm}{\end{itlemma}} 
\newenvironment{remark}{\begin{itremark}\rm}{\end{itremark}} 
\newenvironment{remarks}{\begin{itremarks} \rm}{\end{itremarks}}
\newenvironment{corollary}{\begin{itcorollary}\rm}{\end{itcorollary}}
\newenvironment{proposition}{\begin{itproposition}\rm}{\end{itproposition}}
\newenvironment{definition}{\begin{itdefinition}\rm}{\end{itdefinition}}
\newenvironment{example}{\begin{itexample}\rm}{\end{itexample}}
\newenvironment{fact}{\noindent {\em Fact}. \ \ }{\hfill \medskip}
\newenvironment{proof}{\noindent {\em Proof}.\ \
}{\hspace*{\fill}$\Box$\medskip}
\newenvironment{claim}{\noindent {\em Claim}. \ \ }{\hfill \medskip}
\newcommand{\be}[1]{\begin{equation}\label{#1}}
\newcommand{\ee}{\end{equation}}
\newcommand{\bl}[1]{\begin{lemma}\label{#1}}
\newcommand{\br}[1]{\begin{remark}\label{#1}}
\newcommand{\brs}[1]{\begin{remarks}\label{#1}}
\newcommand{\bt}[1]{\begin{theorem}\label{#1}}
\newcommand{\bd}[1]{\begin{definition}\label{#1}}
\newcommand{\bp}[1]{\begin{proposition}\label{#1}}
\newcommand{\bc}[1]{\begin{corollary}\label{#1}}
\newcommand{\bfact}[1]{\begin{fact}\label{#1}}
\newcommand{\bex}[1]{\begin{example}\label{#1}}
\newcommand{\ec}{\end{corollary}}
\newcommand{\efact}{\end{fact}}
\newcommand{\eex}{\end{example}}
\newcommand{\el}{\end{lemma}}
\newcommand{\er}{\end{remark}}
\newcommand{\ers}{\end{remarks}}
\newcommand{\et}{\end{theorem}}
\newcommand{\ed}{\end{definition}}
\newcommand{\ep}{\end{proposition}}
\newcommand{\epr}{\end{proof}}
\newcommand{\bpr}{\begin{proof}}
\newcommand{\bcl}{\begin{claim}}
\newcommand{\ecl}{\end{claim}}

\newcommand{\bi}{\begin{itemize}}
\newcommand{\ei}{\end{itemize}}
\newcommand{\ben}{\begin{enumerate}}
\newcommand{\een}{\end{enumerate}}
\newcommand{\text}[1]{\hbox{\rm \ #1\ \/}}

\title{GENERAL METHODS TO CONTROL RIGHT-INVARIANT
SYSTEMS ON COMPACT LIE GROUPS AND MULTILEVEL QUANTUM SYSTEMS}

\vs

\vs

\author{Domenico D'Alessandro\thanks{Department of Mathematics, Iowa State
University, Ames, Iowa, U.S.A.\ \ Electronic address:
daless@iastate.edu}}

\begin{document}

\maketitle

 \vs

\begin{abstract}

For a right-invariant system on a compact Lie group $G$, I present
two methods to design a control to drive the state from the identity
to any element of the group. The first method, under appropriate
assumptions, achieves {\it exact} control to the target but requires
estimation of the `size' of a neighborhood of the identity in $G$.
The second method, does not involve any mathematical difficulty, and
obtains control to a desired  target {\it with arbitrary accuracy}.
A third method is then given  combining the main ideas of the
previous methods.  This  is also very simple in its formulation and
turns  out to be generically more efficient as illustrated by one of
the examples we consider.

The methods described in the paper provide arbitrary constructive
control for any right-invariant system on a compact Lie group. I
 give examples  including closed multilevel quantum systems and lossless
electrical networks. In particular, the results can be applied to
the coherent control of general multilevel quantum systems.

\end{abstract}

\newpage

\section{The Lie algebra rank condition of geometric control theory}
\label{intro} Consider a control system of the form \be{genconsys}
\dot x=f(x,u),  \ee where $x$ is the state varying on a compact Lie
group and $u$ the control. The system is said to be {\it right
invariant} if, denoting by $x(t,u,s)$ the solution of
(\ref{genconsys}) corresponding to initial condition $s$ and control
function $u$, we have \be{r1} x(t,u,s)= x(t,u,{\bf 1})\circ s, \ee
where ${\bf 1}$ denotes the identity of the group and $\circ$ is the
multiplication of the group.  To be concrete, we shall consider the
case of matrix groups where the group operation is the standard
matrix multiplication, with particular attention to subgroups of
$SU(n)$, given the potential application to quantum systems. In
particular, we shall consider systems of the form \be{d} \dot
X=A(u)X, \qquad X(0)={\bf 1},  \ee where ${\bf 1}$ is the identity
matrix and the matrix $A(u)$ is in the Lie algebra associated with
$G$ for every value of the control $u$. This equation   models many
systems of interest. In particular closed (i.e., not interacting
with the environment)  finite dimensional quantum systems which are
coherently controlled (i.e., through a variation of their
Hamiltonian) are modeled this way. In this case, equation (\ref{d})
is Schr\"odinger equation. We refer to \cite{Mikobook} and
references therein  for several examples and introductory notions on
Lie groups and Lie algebras in the context of quantum control.

If we restrict ourselves to piecewise constant controls, the problem
of control for systems (\ref{d})  can be described as follows.
Assume that we have a linearly independent set of matrices
\be{effematrices} {\cal F}:=\{ A_1,\ldots, A_m \}. \ee To each of
them there corresponds a semigroup \be{semigruppi} {\cal
S}_j:=\{e^{A_j t} | t\geq 0\}, \qquad j=1,\ldots,m.  \ee The problem
of control to a matrix $X_f$ is to choose $N$ elements $X_l$,
$l=1,\ldots,N$  $ \in {\cal S}_j$, for some $j=1,\ldots,m$, such
that  $\prod_{l=1}^N X_l=X_f$. If such elements exist $X_f$ is said
to be {\it reachable}.  The question of the set of reachable
matrices is a standard one in geometric control theory. The result
in the following Theorem \ref{LARC}, known as the {\it Lie algebra
rank condition}, is classical \cite{JS} and provides the answer for
compact Lie groups.

Let ${\cal L}$ be the Lie algebra generated by the elements in
${\cal F}$ defined as the smallest Lie algebra containing ${\cal F}$
and denote by $e^{\cal L}$ the connected Lie group associated with
${\cal L}$. We shall call  ${\cal L}$ the {\it dynamical Lie
algebra} associated to the system.

\begin{theorem}\label{LARC} \cite{JS} Consider the Lie group
$e^{\cal L}$ and assume it is compact. Then, the  set of reachable
values for $X$ in (\ref{d})  is equal to $e^{\cal L}$.
\end{theorem}

This result has been elaborated upon in several papers and applied
to quantum mechanical systems (cf. \cite{AlbeMiko}, \cite{Mikobook},
\cite{Tarn}, \cite{Rama1}). In particular, in the case of (closed)
quantum mechanical systems ${\cal L}$ is a subalgebra of the unitary
Lie algebra $u(n)$ and, as such, can be written as the direct sum of
an Abelian subalgebra and a semisimple subalgebra to which there
corresponds a compact Lie group.  That is, modulo  an Abelian
subgroup which commutes with all of $e^{\cal L}$, $e^{\cal L}$ is
compact (cf. \cite{MikoLiealg} and \cite{Tannor}). In particular,
$e^{\cal L}$ is compact if ${\cal L}=u(n)$ or ${\cal L}=su(n)$ in
which case, the system is called controllable and $e^{\cal L}$ is
the group of unitary matrices $U(n)$ or special unitary matrices
$SU(n)$, respectively.

The original proof given in \cite{JS} is not constructive, i.e., in
our setting, it does not show how to alternate elements in the
semigroups ${\cal S}_j$ in (\ref{semigruppi}) to obtain a given
target $X_f \in e^{\cal L}$. We show how to obtain this in two ways
in the following two sections. The main ideas are then combined in a
third method in section \ref{M3}. The first method, described in
section \ref{M1}, achieves exact control if the subgroups
corresponding to the semigroups in (\ref{semigruppi}), i.e.,
\be{subgruppi} \tilde {\cal S}_j:=\{ e^{A_jt}|t \in \RR\} , \qquad
j=1,\ldots,m,  \ee are closed. Otherwise it obtains control with
arbitrary accuracy as it follows from Proposition \ref{PJS} and
Remark \ref{nt} below. This proposition allows us to replace an
exponential of the form $e^{A t}$ with $t <0$ with an exponential of
the form $e^{A t}$ with $t
>0$ which approximates it with arbitrary accuracy. This result will
be utilized for  the following two methods as well.

\section{Method 1: Exact constructive  controllability}
\label{M1}

 The method we are going to describe is a consequence of the proof
of the Lie algebra rank condition, Theorem \ref{LARC}, given in
\cite{Mikobook} and the result on uniform finite generation of
compact Lie groups given in \cite{UFG}. Let $X_f \in e^{\cal L}$ be
the target state. We want to show a way to obtain $X_f$ as a product
of elements in (\ref{semigruppi}), if not exactly,  at least, with
arbitrary accuracy. We are first going to relax the problem by
allowing the use of elements in the subgroups $(\ref{subgruppi})$
rather than only elements of the semigroups (\ref{semigruppi}). We
shall show later how to overcome this problem (see Proposition
\ref{PJS} and Remark \ref{nt}).

\vs

Since $e^{\cal L}$ is compact the exponential map is surjective,
that is, there exists a matrix $A \in {\cal L}$ such that
$e^{A}=X_f$, for every $X_f$.\footnote{See, e.g., \cite{Knapp} and
\cite{Mosko} for a study on the generalization of this result.See
also \cite{HornJohnsonT} (Theorem 6.4.15) for the theorem on
existence of the logarithm of a matrix.}  This also implies that,
given any neighborhood $K$ of the identity in $e^{\cal L}$, we can
choose an integer $M$ sufficiently large such that
$e^{\frac{A}{M}}=X_f^{\frac{1}{M}} \in K$. Now, assume first that
${\cal F}$ is a basis for  ${\cal L}$, that is, no Lie bracket is
necessary to obtain a basis of ${\cal L}$. This implies that, by
varying $t_1,\ldots,t_m$ in a neighborhood of the origin in $\RR^m$,
$K:=\{ X=e^{A_s t_s} e^{A_{m-1} t_{m-1}} \cdots e^{A_1 t_1}|
t_1,\ldots,t_m \in \RR\}$, gives a neighborhood of the identity in
$e^{\cal L}$ and, in particular, it contains $e^{\frac{A}{M}}$ for
sufficiently large $M$. That is, we can find real values $\bar
t_1,\ldots,\bar t_m$ such that \be{poi} e^{\frac{A}{M}}=e^{A_m \bar
t_m} e^{A_{m-1} \bar t_{m-1}} \cdots e^{A_1 \bar t_1}.
 \ee
Therefore, by  using elements from the subgroups (\ref{subgruppi})
we can obtain $e^{\frac{A}{M}}$. Now assume ${\cal F}$ is not a
basis for ${\cal L}$.  Since ${\cal F}:=\{ A_1, \ldots, A_m \}$
generates all of ${\cal L}$, there exist two values $1 \leq k, l
\leq m$ such that the commutator $[A_l,A_k]$ is linearly independent
of $\{A_1, \ldots, A_m\}$. This implies that there exists a value $t
\in \RR$ such that $F:=e^{A_lt} A_k e^{-A_lt}$ is also linearly
independent. To see this, assume it is not true and  write $e^{A_l
t} A_k e^{-A_lt}$ as \be{lineacomb} e^{A_l t} A_k
e^{-A_lt}=\sum_{j=1}^m a_j(t) A_j,  \ee for every $t$. Taking the
derivative with respect to $t$ at $t=0$, gives
$[A_l,A_k]=\sum_{j=1}^m \dot a_j(0)A_j$, which contradicts the fact
that $[A_l,A_k]$ is linearly independent of $\{A_1,\ldots,A_m\}$.
Let $\bar t$ be such that \be{effe} F:= e^{A_l \bar t} A_k e^{-A_l
\bar t}. \ee  We can add $F$ to $\{A_1, \ldots, A_m\}$ and still
have a linearly independent set. Moreover, we can express every
exponential $e^{Ft}$ in terms of exponentials of $A_l$ and $A_k$
since $e^{Ft}=e^{A_l \bar t} e^{A_k t} e^{-A_l \bar t}$. Define
$A_{m+1}:=F$. If $\{A_1,\ldots,A_m,A_{m+1}\}$ is a basis of ${\cal
L}$ then we can proceed as above and obtain a neighborhood of the
identity in $e^{\cal L}$ by varying $ \{ t_1,\ldots t_{m+1} \} \in
\RR^{m+1}$. Such a neighborhood is given by  $K:=\{
\prod_{j=1}^{m+1} e^{A_j t_j}| t_1,\ldots,t_{m+1} \in \RR\}$. If
that is not the case, then we observe that $\{A_1,\ldots,A_{m+1} \}$
is still a set of generators for ${\cal L}$ and, as above, there
must exist two elements $A_k$ and $A_l$ in $\{A_1,\ldots,A_{m+1}
\}$, such that $[A_k,A_l]$ is linearly independent of
$\{A_1,\ldots,A_{m+1} \}$ and therefore for some $\bar t$,
$A_{m+2}:=e^{A_l \bar t} A_k  e^{-A_l \bar t}$ is linearly
independent of $\{ A_1,\ldots,A_{m+1}\}$. The exponential
$e^{A_{m+2}t}$ again can be expressed in terms of exponentials of $
A_1,\ldots,A_{m+1}$ and therefore in terms of exponentials of $
A_1,\ldots,A_{m}$. Proceeding this way, one finds $\dim ({\cal
L})-m$ new matrices, $\{A_{m+1}, A_{m+2}, \ldots, A_{\dim({\cal
L})}\}$ which together with $\{A_1,\ldots,A_m\}$ form a basis for
${\cal L}$. By taking $\prod_{j=1}^{\dim({\cal L})} e^{A_j t_j}$
with $t_j \in \RR$, $j=1,\ldots,\dim({\cal L})$, we obtain all the
elements in a neighborhood of the identity and in particular
$e^{\frac{A}{M}}$. Repeating the  sequence $M$ times we obtain
$e^{A}$.

\vs

In the expression of $e^{\frac{A}{M}}$ and therefore in the
expression of $e^{A}$, there will be some exponentials with negative
$t$, i.e., some elements in the subgroups (\ref{subgruppi}) which
are (possibly) not in the semigroups (\ref{semigruppi}). There are
ways to minimize the number of these elements in the full product,
for example by placing together matrices which come from similarity
transformations with the same matrix so as to have cancelations of
the type $e^{A_j t_1}e^{-A_j t_2}=e^{A_j(t_1-t_2)}$. Also, in many
cases, the orbits $\{ e^{A_jt}|t \in \RR\}$   are periodic (closed),
which allows us to assume all the $\bar t_j$'s positive, without
loss of generality. However, if this is not the case we can use the
following fact.

\bp{PJS} Let $e^{-B|t|}$ an element of a compact Lie group $e^{\cal
L}$. For every $\epsilon >0$ there exists a $\bar t >0$ such
that\footnote{Whenever we do specific computations involving norms
of matrices we use the Frobenius norm
$\|A\|:=\sqrt{Trace(AA^\dagger)}$.} \be{lko} \| e^{-B|t|}-e^{B \bar
t} \| < \epsilon. \ee \ep \bpr Consider $e^{-B|t|}$ and the sequence
$e^{nB|t|}$, which by compactness of $e^{\cal L}$ has a converging
subsequence $e^{n(k)B|t|}$. We have $\lim_{k \rightarrow \infty}
e^{(n(k+1)-n(k)-1)B|t|}=e^{-B|t|}$. Therefore there is $\bar k$ such
that $ \|e^{(n(\bar k+1)-n(\bar k)-1)B|t|} - e^{-B|t|}\| <
\epsilon$, and the proposition holds with $\bar t= (n(\bar
k+1)-n(\bar k)-1)|t|$. \epr

\vs

\br{nt} The proof given above follows the one given in \cite{JS}. A
different, more concrete, proof can be given for Lie subgroups of
$U(n)$, which is the case that interests us the most. In that case,
using the Frobenius norm of matrices, we have \be{Frob} \left\|
e^{B\bar t}- e^{-B |t|}  \right\|=\sqrt{2} \sqrt{n-\sum_{j=1}^n
cos(\omega_j(\bar t+|t|))},  \ee where $i\omega_j$, $j=1,\ldots,n$
are the eigenvalues (possibly repeated) of $B$. If we can choose
$\bar t >0 $ so that \be{Dirich} \left[1-\cos(\omega_j(\bar
t+|t|))\right]< \frac{\epsilon^2}{2n},  \ee for every
$j=1,\ldots,n$, then (\ref{lko}) is certainly satisfied. If
$g:=\arccos\left(1-\frac{\epsilon^2}{2n}\right)$, then, we satisfy
condition (\ref{Dirich}) if we are able to find $\bar t$ and
integers $m_j$, $j=1,\ldots,n$ such that \be{hui} \left|
\omega_j(\bar t +|t|)-2 \pi m_j  \right| <g. \ee However, according
to Dirichlet's approximation theorem (see, e.g., \cite{Cassels}),
given a natural number $N$ and $n$ reals $\alpha_1,\ldots,
\alpha_n$, we can find positive integers $a, b_1,\ldots,b_n$, with
$1 \leq a \leq N^n$ so that $|\alpha_j a-b_j| < \frac{1}{N}$. This
result can be applied to satisfy condition (\ref{hui}) identifying
$\alpha_j$ with $\frac{\omega_j |t|}{2 \pi}$ and choosing
$\frac{1}{N} < \frac{g}{2\pi}$ and choosing $m_j=b_j$ and $\bar t$
so that $\frac{\bar t +|t|}{|t|}=a$. Notice that since $a \geq 1$,
$\bar t \geq 0$ as desired. For the problem to find $a$ and $b_j$'s,
there are several algorithms in the literature (cf. \cite{Germans}
and \cite{Continuedfractions}). Notice, in any case, that we are
only interested in $a$, which determines $\bar t$, and since $a$ is
bounded from above by $N^n$, it can be always found, in principle,
by exhaustive search.  \er

\vs

\vs

We can summarize the given  method as follows:

\begin{enumerate}

\item Given ${\cal F}:=\{A_1,\ldots,A_m \}$ find, via similarity transformations,
$\dim{\cal L}-m$ more matrices $\{A_{m+1},\ldots,A_{\dim({\cal
L})}\}$ so that $\{A_1,\ldots,A_{\dim{({\cal L})}} \}$ is a basis
for ${\cal L}$.

\item Take the (principal) logarithm of $X_f$, $A$, so that
$e^{A}=X_f$.

\item Find $M$ (sufficiently large) and $t_1, \ldots, t_{\dim({\cal L})}$,
so that \be{basicprod} e^{\frac{A}{M}}=\prod_{j=1}^{\dim({\cal L})}
e^{A_j t_j}. \ee
 Then
$X_f=e^{A}=\left(\prod_{j=1}^{\dim({\cal L})} e^{A_j t_j}\right)^M$.

\item Replace the exponentials of the matrices
$A_{m+1},\ldots,A_{\dim({\cal L})}$ with expressions involving the
exponentials of $\{ A_1, \ldots, A_m\}$ as obtained from step 1.

\item Replace every exponential $e^{Bt}$, $(B \in {\cal F})$ involving negative $t$ with
its approximation involving positive $t$. This can be obtained with
arbitrary accuracy according to Proposition \ref{PJS} and Remark
\ref{nt}.

\end{enumerate}

\vs

\br{rem1} In the above procedure, step 3. is decidedly the most
difficult one since it requires the solution of nonlinear equations
involving the exponentials of matrices. The solution is guaranteed
to exist for $M$ sufficiently large.  This task is obviously easier
for low dimensional systems. It must be remarked however that there
is some flexibility in the choice of the matrices $A_{m+1}, \ldots,
A_{\dim({\cal L})}$, because of the choice of the pair $A_k,A_l$ and
of the times $\bar t$ (cf. (\ref{effe})).  We can use this
flexibility to make these matrices as simple as possible (e.g.,
block diagonal, sparse, etc.) so that calculating the exponential is
easier. Another type of flexibility, which may be used in
calculations, is the fact that the way exponentials are arranged in
(\ref{basicprod}) is arbitrary. Any different order will give a
neighborhood of the identity also. The methods described in the
following two sections do not present this problem.

\er

\br{rem2} The last step of the method can be achieved exactly (i.e.,
without involving an approximation) if the orbit associated with the
given matrices ${\cal F}:=\{A_1,\ldots,A_m\}$ are periodic. In this
respect, notice that, if this is the case, all the other matrices
obtained by the method also have associated periodic orbits (their
eigenvalues are the same as the ones of the original matrices).
Therefore, for a given matrix $B$, and  negative $\bar t$, we can
choose a positive $t$, such that $e^{Bt}=e^{B\bar t}$. \er

\br{rem3} \cite{UFG} It is interesting to give an upper  bound to
the number of exponentials involved in obtaining a neighborhood of
the identity according to the described method. Let us assume that,
at every step,  we only produce one new linearly independent matrix.
For the given matrices $\{A_1, \ldots, A_m\}$, we need only one
exponential, but for the matrix obtained at step 1 we need three
exponentials. In general, at step $j$, $j \geq 2$, the worst case
scenario is when we combine a matrix obtained at step $j-1$ (giving
the similarity transformation ($A_l$ in (\ref{effe})), which
requires $d_{j-1}$ exponentials, with a matrix obtained at step
$j-2$, which requires $d_{j-2}$ exponentials. The total number of
exponentials at step $j$ is therefore $d_j=2d_{j-1}+ d_{j-2}$.
Therefore having defined recursively the numbers $d_j$ as \be{recor}
d_0=1,\qquad d_1=3, \qquad d_j=2d_{j-1}+d_{j-2},  \ee the number of
exponentials required is \be{numberexpo} md_0+\sum_{j=1}^{\dim{\cal
L}-m} d_j. \ee \er

\subsection{Example}

We illustrate this method with a simple example of the quantum
control of a two level system, i.e., a control problem on $SU(2)$,
which is compact. Recall the definition of the Pauli matrices

\be{PauliMat} \sigma_x:=\pmatrix{0 & 1 \cr 1 & 0}, \qquad
\sigma_y:=\pmatrix{0 & i \cr -i & 0}, \qquad \sigma_z:=\pmatrix{1 &
0 \cr 0 & -1}.  \ee

Let ${\cal F}:=\{A_1, A_2\}$, with $A_1:=i\sigma_z$ and $A_2:=i
(\sigma_x+ \sigma_y)$. Calculate $e^{A_1 \bar t}A_2 e^{-A_1 \bar t}$
which for $\bar t=-\frac{3}{8}\pi$ gives $A_3=-i\sqrt{2} \sigma_y$,
which is linearly independent of $A_1$ and $A_2$, and along with
them it forms a basis of $su(2)$.   A straightforward calculation
gives \be{esponenziali} e^{A_1t_1}=\pmatrix{e^{it_1} & 0 \cr 0 &
e^{-it_1}}, \quad e^{A_2t_2}=\pmatrix{\cos(\sqrt{2}t_2) & e^{i
\frac{3\pi}{4}} \sin(\sqrt{2}t_2) \cr - e^{-i \frac{3\pi}{4}}
\sin(\sqrt{2}t_2) & \cos(\sqrt{2}t_2)} \ee
$$
e^{A_3t_3}=\pmatrix{\cos(\sqrt{2}t_3) & \sin(\sqrt{2}t_3) \cr -
\sin(\sqrt{2}t_3) & \cos(\sqrt{2}t_3)}.
$$
and the  the set \be{sdlfirst} S_{1,2,3}:=\{e^{A_1 t_1} e^{A_2 t_2}
e^{A_3 t_3}|t_1,t_2,t_3 \in \RR\}, \ee covers a neighborhood of the
identity in $SU(2)$. Assume now our target state $X_f$ is
\be{targets} X_f:=\pmatrix{\frac{1}{\sqrt{2}} &
i\frac{1}{\sqrt{2}}\cr i \frac{1}{\sqrt{2}} & \frac{1}{\sqrt{2}}}.
\ee We first try to see if $X_f$ is in the set $S_{1,2,3}$ in
(\ref{sdlfirst}). Therefore we must be able to choose $t_1$ and
$t_3$ so that $P:=e^{-A_1 t_1} X_f e^{-A_3 t_3}$ has the form
$e^{A_2 t_2}$. This means in particular that the difference between
the phases of the $P_{1,2}$ element and $P_{1,1}$ elements in $P$ is
$\frac{3 \pi}{4}$. As a straightforward calculation shows,
$P_{1,2}P_{1,1}^*=\frac{i}{2}$ independently of the choice of $t_1$
and $t_3$. Therefore $X_f \notin S_{1,2,3}$. We replace $X_f$ with
$X_f^{\frac{1}{2}}$. The same calculation shows that, for every
$t_1$, $P_{1,2}P_{1,1}^*=\frac{\sqrt{2}}{2} \sin(2 \sqrt{2}t_3)+i
\frac{\sqrt{2}}{2}$ and, therefore, the choice $t_3:=\frac{3 \pi}{4
\sqrt{2}}$ achieves the desired phase difference. Then, we can
choose $t_1$ to impose that the element $P_{1,1}$ has phase zero (it
is real). This leads to $t_1=\frac{9 \pi}{8}$. With these choices,
we have \be{afterchoices} e^{-A_1 t_1} X_f^{\frac{1}{2}} e^{-A_3
t_3}=\pmatrix{ \frac{1}{\sqrt{2}} &  \frac{1}{\sqrt{2}} e^{i \frac{3
\pi}{4}} \cr -  \frac{1}{\sqrt{2}} e^{-i \frac{3 \pi}{4}} &
\frac{1}{\sqrt{2}}}. \ee Comparing this with $e^{A_2t_2}$ in
(\ref{esponenziali}) leads to the choice
$t_2=\frac{\pi}{4\sqrt{2}}$. With these choices $X_f=\left(e^{A_1
t_1}e^{A_2 t_2} e^{A_3 t_3} \right)^2$.

In terms of the original available matrices,  $A_1$ and $A_2$, we
have \be{hjklo} X_f=\left(e^{A_1 t_1}e^{A_2 t_2} e^{-A_1 \frac{3
\pi}{8}} e^{A_2t_3} e^{A_1 \frac{3 \pi}{8}} \right)^2,  \ee where
$t_1,t_2,t_3$ are the ones found above. The presence  of the
negative `time' $-\frac{3 \pi}{8}$ in the third exponential, does
not pose any problems since the one dimensional subgroup associated
with $A_1$ (as well as any other matrix in $su(2)$) is periodic.

\vs

A similar treatment shows that, had we chosen to work with the set
\be{sdl}
 S_{1,3,2}:=\{e^{A_1 t_1} e^{A_3 t_3} e^{A_2
t_2}|t_1,t_2,t_3 \in \RR\}, \ee we would have achieved $X_f$ with
just three exponentials. This  shows that the order in which the
exponentials are chosen may be important.

\vs

It must be said that for the special case of $SU(2)$ there are many
more techniques which may be preferable to the one advocated here.
For example, since one has available both $i\sigma_z$ and $i
\sigma_y$ one could have applied a simple Euler decomposition. In
general it is also possible, for general target matrices,  to find
the factorization with the minimum number of factors \cite{ioopt}.
Our goal here was to illustrate the method on a simple, easily
computable, case.  We remark that even for large dimensional Lie
groups, one can combine these ideas with Lie group decompositions
for which there exists a large set of tools \cite{Mikobook}.

\section{Method 2: Constructive controllability with arbitrarily
small error} \label{M2}

In this and the following section we illustrate methods which do not
require the solution of nonlinear algebraic equations, such as
(\ref{poi}), but can be implemented with simple linear algebraic
techniques. The algorithms achieve control to the target with
arbitrary small error.

 Reconsider the available set of matrices
${\cal F}$ in (\ref{effematrices}). As before, we relax the
requirement to use only  elements in the semigroups
(\ref{semigruppi}) and use elements in the subgroups
(\ref{subgruppi}). We can then replace elements in the subgroups
with elements in the semigroups as done in the previous section.
 We
start with a definition

\bd{approximable} A matrix $H$ is said to be {\it simulable} with
the set ${\cal F}$ if there exist $r$ continuous, strictly
increasing, functions $f_j$, $j=1,\ldots,r$, with $f_j(0)=0$,
defined in an interval $[0,\epsilon)$, such that \be{AIG}
e^{Hx}=\prod_{j=1}^r e^{L_j f_j(x)}+O(x^{1+\delta}),  \ee for some
matrices $L_j \in {\cal F} \bigcup -{\cal F} $ and\footnote{$-{\cal
F}$ denotes the set $\{-A_1,-A_2,\ldots,-A_m\}$.} a $\delta
>0$. \ed

If a matrix $H$ is simulable,  we can control from the identity to
$e^H$ with the desired accuracy using elements in the subgroups
(\ref{subgruppi}) (and therefore of the semigroups
(\ref{semigruppi})).

\bl{Limitefondamentale} Assume (\ref{AIG}) holds. Then \be{baslim}
\lim_{n \rightarrow \infty} \left( \prod_{j=1}^r e^{L_j
f_j(\frac{1}{n})} \right)^n=e^H \ee \el

\bpr If (\ref{AIG}) holds then \be{polp} \lim_{n \rightarrow \infty}
\left( \prod_{j=1}^r e^{L_j f_j(\frac{1}{n})} \right)^n= \lim_{n
\rightarrow \infty} \left[e^{H \frac{1}{n}}-
O\left(\frac{1}{n^{1+\delta}}  \right) \right]^n.  \ee However, we
have this standard limit in matrix analysis (see,
\cite{HornJohnsonT} section 6.5) \be{standlim} \lim_{n \rightarrow
\infty} \left[e^{H \frac{1}{n}}- O\left( \frac{1}{n^{1+\delta}}
\right) \right]^n =e^H, \ee which proves the lemma. \epr

\vs

From the point of view of constructive controllability, this lemma
says that, for each simulable $H$,  we can put together a product of
exponentials of elements in ${\cal F}$ which, repeated a large
enough number of times, approximates, with arbitrary accuracy,
$e^{H}$.

\vs

\bt{fondafonda} Every $H$ in the dynamical Lie algebra ${\cal L}$ is
simulable.

\et

\br{rem1p} This theorem along with Lemma \ref{Limitefondamentale}
and Proposition \ref{PJS} give an alternative proof of a slightly
weaker form of the Lie algebra rank condition of Theorem \ref{LARC}.
Since $e^{\cal L}$ is compact, for every $X_f$ in $e^{\cal L}$,
there exists an $H \in {\cal L}$ such that $e^{H}=X_f$. Theorem
\ref{fondafonda} and Lemma \ref{Limitefondamentale} say that we can
find a sequence of reachable points converging to $X_f$ for every
$X_f$. Therefore the  set of reachable states  is dense in $e^{\cal
L}$. \er

\br{rem2p} Elaborating on the  proof of the Theorem
\ref{fondafonda}, we  will also show how to choose the elements $L_j
\in {\cal F} \bigcup -{\cal F}$ and the functions $f_j$ in
(\ref{AIG}) so as to make the controllability result constructive.
We shall discuss this after the proof.  \er

\bpr The proof is similar to the one given in \cite{IOQW} in the
context of quantum walks dynamics. In particular, we will show that
the set of simulable elements $H$ is a Lie algebra containing ${\cal
F}$ and this will be sufficient since ${\cal L}$ is the smallest Lie
algebra containing ${\cal F}$, by definition.

First of all, it is clear that every element in ${\cal F}$ is
simulable, since equation (\ref{AIG}) holds with $r=1$ and $O\equiv
0$. Therefore the set of simulable matrices contains ${\cal F}$.

Moreover if $H$ satisfies equation (\ref{AIG}), then we have
\be{AIGinverse} e^{-Hx}= \prod_{j=r}^1e^{-L_j f_j(x)}-
\prod_{j=r}^1e^{-L_j f_j(x)} O(x^{1+\delta}) e^{-Hx}, \ee and by
expanding the exponentials it follows that the last term is also an
$O(x^{1+\delta})$. Therefore $-H$ is also simulable. Moreover, for
$a \geq 0$, (\ref{AIG}) holds for $aH$ with $f_j(x)$ replaced by
$f_j(ax)$ and $O(x^{1+\delta})$ replaced by
$O(a^{1+\delta}x^{1+\delta})=O(x^{1+\delta})$. If (\ref{AIG}) holds
for $H_1$ and $H_2$, i.e., we have \be{H12} e^{H_i
x}=\prod_{j=1}^{r_i} e^{L_j^i f_j^i(x)}+ O_i(x^{1+\delta_i}), \qquad
i=1,2, \ee combining this with \be{somma}
e^{(H_1+H_2)x}+O(x^2)=e^{H_1x} e^{H_2x},  \ee gives\footnote{Here
and elsewhere, we use the notation $O$ for a generic $O$-function
and we use indexes like in $O_1$ and $O_2$ when we want to highlight
a particular $O$-function.}  \be{popg}
e^{(H_1+H_2)x}=\prod_{j=1}^{r_2} e^{L_j^2 f_j^2(x)}
\prod_{j=1}^{r_2} e^{L_j^1 f_j^1(x)}+ O(x^{1+ \delta}), \ee with
$\delta=\min \{ \delta_1,\delta_2,1 \}.$ Therefore, if $H_1$ and
$H_2$ are simulable, so is $H_1 + H_2$. These arguments show that
the set of simulable matrices is a vector space.

To show that it is also a Lie algebra, we have to show that if $H_1$
and $H_2$ are both simulable so is $[H_1,H_2]$. In order to see
that, write (\ref{H12}) in the form \be{H12form}
e^{H_1t}=T_1(t)+O_1(t^{1+\delta_1}), \qquad
e^{H_2t}=T_2(t)+O_2(t^{1+\delta_1}),  \ee i.e., by replacing the
products with the functions $T_1$ and $T_2$. This also gives (cf.
(\ref{AIGinverse})) \be{H12forminverse} e^{-H_1 t}=T_1^{-1}(t)-
T_1^{-1}(t)O_1(t^{1+\delta_1}) e^{-H_1t}, \quad  e^{-H_2
t}=T_2^{-1}(t)- T_2^{-1}(t)O_2(t^{1+\delta_2}) e^{-H_2t}. \ee We use
the exponential formula (see, e.g., \cite{HornJohnsonT} Section 6.5)
\be{expofor} e^{[H_1,H_2]t^2}+O(t^3)=e^{-H_1 t}e^{-H_2t} e^{H_1
t}e^{H_2t}.\ee Using (\ref{H12form}) and (\ref{H12forminverse}) in
(\ref{expofor}), we have \be{pplm} e^{[H_1, H_2]t^2}+O(t^3)= \left(
T_1^{-1} - T_1^{-1} O_1 e^{-H_1t} \right) \left( T_2^{-1}- T_2^{-1}
O_2 e^{-H_2 t} \right) \left( T_1+O_1 \right) \left( T_2 +O_2
\right). \ee Expanding the right hand side, omitting terms that are
clearly $O(t^{\alpha})$, $\alpha > 2$, since they contain the
product of two $O$ functions, we have \be{dfd}
e^{[H_1,H_2]t^2}+O(t^3)=T^{-1}_1 T^{-1}_2 T_1 T_2 +
T_1^{-1}T_2^{-1}T_1 O_2+ T_1^{-1} T_2^{-1} O_1 T_2 \ee $$- T_1^{-1}
T_2^{-1} O_2 e^{-H_2 t} T_1 T_2+ T_1^{-1} O_1 e^{-H_1 t} T_2^{-1}
T_1 T_2 +O(t^\alpha).  $$ Expanding in McLaurin series the functions
multiplying the $O_1$ and $O_2$, we see that the terms corresponding
to the first terms of the expansion cancel, leaving only terms of
the form $O(t^{\beta'})$ with $\beta' >2$. In conclusion,  we have
\be{jhu} e^{[H_1, H_2]t^2}=T_1^{-1}(t)T_2^{-1}(t) T_1(t) T_2(t)+
O(t^\beta), \qquad \beta >2,  \ee and by setting $t=\sqrt{x}$, we
obtain \be{klko} e^{[H_1,
H_2]x}=T_1^{-1}(\sqrt{x})T_2^{-1}(\sqrt{x}) T_1(\sqrt{x})
T_2(\sqrt{x})+ O(x^{\frac{\beta}{2}}), \qquad \beta
>0,
\ee which shows that $[H_1,H_2]$ is simulable as well, and completes
the proof. \epr

\vs

In order to use Lemma \ref{Limitefondamentale} and Theorem
\ref{fondafonda} for control, we need to show, given $H$,  how to
find the matrices $L_j$ in ${\cal F} \bigcup - {\cal F}$ so that
(\ref{AIG}) holds. We first find a basis of ${\cal L}$ by taking
repeated Lie brackets of elements in ${\cal F}$. More precisely, set
\be{D0} {\cal D}_0:={\cal F}, \ee a linearly independent set of
elements of `{\it depth}' $0$ (no Lie bracket necessary), and let
\be{olop} \tilde {\cal D}_1:=[{\cal D}_0, {\cal F}], \ee a set of
elements of depth 1, which are Lie brackets of elements of depth $0$
with elements of ${\cal F}$. From the set $\tilde {\cal D}_1$ we
extract a possibly smaller set ${\cal D}_1$ such that ${\cal D}_0
\bigcup {\cal D}_1$ is a maximal linearly independent set in ${\cal
D}_0 \bigcup \tilde {\cal D}_1$. Proceeding this way, we now
calculate a set of Lie brackets of depth $2$ \be{D2} \tilde {\cal
D}_2:=[{\cal D}_1, {\cal F}],  \ee and extract a subset ${\cal D}_2
\subseteq \tilde {\cal D}_2$ so that ${\cal D}_0 \bigcup {\cal D}_1
\bigcup {\cal D}_2$ is a maximal linearly independent set in ${\cal
D}_0 \bigcup {\cal D}_1 \bigcup \tilde {\cal D}_2$. Proceeding this
way, we obtain a set $\bigcup_{k=0}^r {\cal D}_k$, which spans all
of ${\cal L}$. As a consequence of ${\cal L}$ being finite
dimensional, the procedure will end at some finite depth $r$ after
which we cannot find any new linearly independent matrix. We write,
for $k=0, \ldots, r$, \be{jjkk} {\cal D}_k;=\{ D_{1k},D_{2k},
\ldots, D_{n_kk} \}. \ee We can decompose $H$ as \be{klo}
H=\sum_{k=0}^rH_k, \ee with $H_k$ a linear combination of elements
of depth $k$, that is, \be{AKK} H_k:=\sum_{j=1}^{n_k}
\alpha_{kj}D_{jk}. \ee Now, following the proof of the theorem, we
can write \be{ops} e^{Hx}=\prod_{k=0}^r e^{H_k x} + O(x^{1+
\delta}). \ee Then we can write each of the $e^{H_k x}$ as \be{dfdp}
e^{H_k x}=\prod_{j=1}^{n_k} e^{D_{jk}f_j(x)}+ O(x^{1+\delta_k}), \ee
for some $\delta_k >0$. This is  straightforward for $k=0$ and it
has to be done iteratively for Lie brackets of higher depth
following the procedure indicated in the proof of theorem.
Summarizing the method is as follows:

\begin{enumerate}
\item Find a basis for ${\cal L}$ by repeated Lie brackets of
elements of ${\cal F}$. Let $r$ denote the maximum depth.

\item Expand $H$ as a sum of linear combinations of matrices of
depth $0,1,\ldots$, as in (\ref{klo}), (\ref{AKK}).

\item For each of these linear combinations approximate the
exponential  with a product of exponentials involving elements in
the basis according to the proof of theorem \ref{fondafonda}. In
particular the rules to obtain the approximating products are as
follows (see proof of theorem \ref{fondafonda}).
\begin{enumerate}

\item If $A \in {\cal F} \cup -{\cal F}$, then the associated
product is $T(x)=e^{Ax}$ (only one factor).

\item If $T(x)$ is the product associated with $A$, then $T^{-1}(x)$ is
the product associated with $-A$.

\item If $T(x)$ is the product associated with $A$, then $T(a
x)$ is the product associated with $a A$ for any  $a \geq 0$.

\item If $T_A(x)$ and $T_B(x)$ are the products associated with $A$
and $B$ respectively, then $T_A(x)T_B(x)$ is the product associated
with $A+B$.

\item If $T_A(x)$ and $T_B(x)$ are the products associated with $A$
and $B$ respectively, then
$T_A^{-1}(\sqrt{x})T_B^{-1}(\sqrt{x})T_A(\sqrt{x})T_B(\sqrt{x})$ is
the product associated with $[A,B]$.

\end{enumerate}

\item Combine all the products in a unique product approximating
$e^{Hx}$, which contains only exponentials of elements in ${\cal F}$
and $-{\cal F}$. By repeating this product for $x=\frac{1}{n}$ a
large number of times $n$ we obtain a matrix arbitrarily close to
$e^{H}$.

\item Replace every exponential $e^{At}$  with $A \in {\cal F}$ and $t <0$
in the approximating product  with an approximating exponential of
the form $e^{A \bar t}$ with $\bar t >0$, according to proposition
\ref{PJS} and remark \ref{nt}.

\end{enumerate}

\subsection{Example}
\label{subsec}

We illustrate the previous procedure with an example taken from the
theory of electrical networks. In particular, we consider the LC
switching network in \cite{Wood} (see also \cite{Ramak}) whose
dynamical equation is given by \be{dyneq} \dot x=\pmatrix{0 & -\nu &
0 & 0 \cr \nu & 0 & 0 & 0 \cr 0 & 0 & 0 & -\beta \cr 0 & 0 & \beta &
0} x +  \pmatrix{0 & 0 & 0 & \gamma \cr 0 & 0 & \delta & 0 \cr 0 &
-\delta & 0 & 0 \cr - \gamma & 0 & 0 & 0 }x u(t), \ee where $\nu$,
$\beta$, $\gamma$ and $\delta$ are positive parameters depending the
inductances and capacitances of the electrical network. The vector
$x$ represents voltages and currents in the network and $u$ is a
switching control variable which takes values in $\{ 0,1 \}$. To
make the discussion  concrete,  we choose the parameters $\nu=1$,
$\beta=3$, $\gamma=1$ and $\delta=2$, so that the set of available
matrices is \be{calfexample} {\cal F}:=\left\{A_1:=\pmatrix{0 & -1 &
0 & 1 \cr 1 & 0 & 2 & 0 \cr 0 & -2 & 0 & -3 \cr -1 & 0 & 3 &
0},\quad  A_2:= \pmatrix{0 & -1 & 0 &  0 \cr 1 & 0 & 0 & 0 \cr 0 & 0
& 0 & -3 \cr 0 & 0 & 3 & 0} \right\}. \ee The solution of
(\ref{dyneq}) is \be{hjd} x(t)=X(t) x(0),  \ee where $X=X(t)$ is the
solution of the matrix equation \be{newad} \dot X=A(u) X, \quad
X(0)={\bf 1}, \quad A(1)=A_1, \quad A(0)=A_2.   \ee

Let us use the notation $E_{jk}$ for the skew-symmetric $ 4 \times 4
$ matrix which has all the entries equal to zero except for the
$(jk)$-th and $(kj)$-th ($1\leq j<k \leq 4$) which are equal to $1$
and $-1$, respectively. Therefore, we can write \be{fdg}
A_1=-E_{12}+E_{14}+2E_{23}-3E_{34}, \qquad A_2=-E_{12}-3E_{34}.  \ee
By calculating Lie brackets, at depth 1,  we obtain \be{A3}
A_3:=[A_2,A_1]=-5 E_{12}+7 E_{24},  \ee at depth 2 \be{A4A5}
A_4:=[A_3, A_1]=17 E_{12}+22 E_{14}+26E_{23}+19 E_{34}, \text{ and }
A_5:=[A_3, A_2]=22 E_{14}+ 26 E_{23}.  \ee At depth 3, we obtain
\be{A6} A_6:=[A_4,A_1]= 145 E_{13}-155 E_{24}. \ee As the matrices
$\{A_l \}$, $l=1,\ldots,6$, are linearly independent, they span all
of $so(4)$ and system (\ref{newad}) varies on the Lie group $SO(4)$,
a compact Lie group.

Let us denote by $T_j=T_j(x)$ the products approximating $e^{A_j
x}$, $j=1,\ldots,6$,  and let us assume that the control problem is
to transfer the state $[0,0,0,1]^T$ to $[1,0,0,0]^T$. We choose to
drive the transition matrix $X$ in (\ref{newad})   to the value
\be{pog} e^{A_5 \frac{\pi}{44}}=\pmatrix{0 & 0 & 0 & 1 \cr 0 &
\cos(\frac{13 \pi}{22}) & \sin (\frac{13 \pi}{22}) & 0 \cr 0 & -
\sin (\frac{13 \pi}{22})  &  \cos(\frac{13 \pi}{22}) & 0\cr -1 & 0 &
0 & 0}.\ee We proceed using the composition rules illustrated in
(a)-(e) above. Since $A_5=[A_3,A_2]$, we have \be{lol}
T_5(x)=T_3^{-1}(\sqrt{x}) T_2^{-1}(\sqrt{x}) T_3(\sqrt{x})
T_2(\sqrt{x}). \ee Moreover, since $A_3=[A_2,A_1]$ we have \be{3lol}
T_3(x)=T_2^{-1}(\sqrt{x}) T_1^{-1}(\sqrt{x}) T_2(\sqrt{x})
T_1(\sqrt{x}), \ee and replacing into (\ref{lol}), we obtain
\be{lolll} T_5(x)= \ee $$T_1^{-1}(\root {4} \of {x}) T_2^{-1}(\root
{4} \of {x}) T_1(\root {4} \of {x}) T_2(\root {4} \of {x})
T_2^{-1}(\sqrt{x}) T_2^{-1}(\root {4} \of {x}) T_1^{-1}(\root {4}
\of {x}) T_2(\root {4} \of {x}) T_1(\root {4} \of {x})
T_2(\sqrt{x}).$$ The product approximating $e^{A_5 \frac{\pi}{44}t}$
is $T_5(\frac{\pi}{44}t)$ which we can express in terms of
exponentials of $A_1$ and $A_2$ only by replacing $T_1$ and $T_2$
(and $T_1^{-1}$ and $T_2^{-1}$) according to the rules in (a) and
(b) above. In conclusion, we have \be{oood} T_5\left(\frac{\pi}{44}t
\right)= e^{-A_1(\frac{\pi}{44}t)^{\frac{1}{4}}}
e^{-A_2(\frac{\pi}{44}t)^{\frac{1}{4}}}
e^{A_1(\frac{\pi}{44}t)^{\frac{1}{4}}}
e^{A_2(\frac{\pi}{44}t)^{\frac{1}{4}}}
e^{-A_2(\frac{\pi}{44}t)^{\frac{1}{2}}}\times \ee
$$
e^{-A_2(\frac{\pi}{44}t)^{\frac{1}{4}}}
e^{-A_1(\frac{\pi}{44}t)^{\frac{1}{4}}}
e^{A_2(\frac{\pi}{44}t)^{\frac{1}{4}}}
e^{A_1(\frac{\pi}{44}t)^{\frac{1}{4}}}
e^{A_2(\frac{\pi}{44}t)^{\frac{1}{2}}}. $$ We numerically calculated
the error \be{errore} Err^2(n)=\left\| e^{A_5
\frac{\pi}{44}}-\left[T_5\left(\frac{\pi}{44}\frac{1}{n}\right)\right]^n
\right\|^2= 8-2Tr \left[
\left(T_5\left(\frac{\pi}{44}\frac{1}{n}\right)\right)^n  e^{A_5^T
\frac{\pi}{44}}   \right], \ee for various values of $n$ and the
behavior of the Error as a function of the number of iterations $n$
is reported in Table \ref{tavola}. The error goes to zero as
predicted by the above treatment. In a log-log scale the behavior is
essentially linear.


\begin{table}[ht]
\caption{Results of numerical experiments for the method in section
\ref{M2}.} \vs \centering
\begin{tabular}{c c}
\hline \hline number of iterations $n$ & Error $Err$\\
\hline 2 & 3.1531 \\
\hline 10 & 2.3964 \\
\hline 20 & 2.0500 \\
\hline  30 & 1.8604 \\
\hline 100 & 1.3761 \\
\hline 500 & 0.9089 \\
\hline 1000 & 0.7599 \\
\hline 5000 & 0.5022 \\
\hline 50000 & 0.2791\\
\hline 100000 & 0.2341 \\
\hline 500000 & 0.1558 \\
\hline 1000000 & 0.1301 \\
\hline 5000000 & 0.0873 \\
\hline 10000000 & 0.0733 \\
\hline 50000000 & 0.0490 \\
\hline 100000000 & 0.0411 \\
\hline
 \label{tavola}
\end{tabular}
\end{table}

\vs

To conclude the example we have to solve the problem that negative
times are not allowed and therefore we have to replace terms of the
form $e^{-A_1 x}$ and $e^{-A_2 x}$, with $x >0$ in the expression of
$T_5$ with approximations  of the form $e^{A_1 x}$ and $e^{A_2 x}$,
respectively. In the case of $A_2$, since $\{e^{A_2 t}| t \in \RR\}$
is periodic we can always find $x_1 >0$ such that $e^{-A_2 x}=e^{A_2
x_1}$ for every $x$, and we can simply replace the exponential with
$x$ with the exponential with $x_1$ in $T_5$, without changing the
error. For the exponentials of $A_1$ however we need to find an
approximation and this is always possible with arbitrary accuracy
according to proposition \ref{PJS} and remark \ref{nt}.

\vs

To be concrete let us assume that the maximum error we can tolerate
is $0.4$. From Table 1, we choose $n=10^{5}$. Fix $x:=\frac{\pi}{44}
{10^{-5}}$.  We have (cf. (\ref{errore}) and Table \ref{tavola})
\be{hhff} Err(10^5)=\left\| e^{A_5 \frac{\pi}{44}}-\left( T_5(x)
\right)^{10^5} \right\| < 0.235.  \ee Let $\tilde T_5$ be the
approximation of $T_5$ in (\ref{oood}) where we only use positive
values in  the exponentials, appropriately replacing the
exponentials of $A_1$. In particular, by rewriting $T_5(x)$ in
(\ref{oood}) as \be{rewrit} T_5(x)=e^{-A_1 x^{\frac{1}{4}}} \Pi_1(x)
e^{-A_1 x^{\frac{1}{4}}} \Pi_2(x), \ee with $\Pi_1(x)=e^{-A_2
x^{\frac{1}{4}}} e^{A_1 x^{\frac{1}{4}}} e^{A_2 x^{\frac{1}{4}}}
e^{-A_2 x^{\frac{1}{2}}} e^{-A_2 x^{\frac{1}{4}}}$ and
$\Pi_2(x)=e^{A_2 x^{\frac{1}{4}}} e^{A_1 x^{\frac{1}{4}}} e^{A_2
x^{\frac{1}{2}}}$, we have \be{tildeT5} \tilde T_5:=\tilde
T_5(x_1,x)=e^{A_1 x_1} \Pi_1(x) e^{A_1 x_1} \Pi_2(x).  \ee Therefore
the actual error $\tilde Err$ is given by \be{actualerr} \tilde
Err=\left\| e^{A_5 \frac{\pi}{44}} -\left[\tilde
T_5(x,x_1)\right]^{10^5} \right\| \leq \left\| e^{A_5
\frac{\pi}{44}} -\left[T_5(x) \right]^{10^5} \right\|+ \left\|
\left[T_5(x) \right]^{10^5}- \left[\tilde T_5(x,x_1) \right]^{10^5}
\right\| \ee
$$<0.235 + \left\| \left[T_5(x)\right]^{10^5}-
\left[\tilde T_5(x,x_1)\right]^{10^5} \right\|,
$$ where we used (\ref{hhff}). Using the formula for  $A$
and $B$ unitary matrices\footnote{This formula  is proved by writing
$A^n-B^n=\sum_{k=1}^n A^{n-k}(A-B)B^{k-1}$, which gives
$$
\left\| A^n-B^n \right\| \leq \sum_{k=1}^n \left\|
A^{n-k}(A-B)B^{k-1} \right\|=n \| A- B \|,
$$
since multiplication (right or left) by a unitary matrix does not
modify the Frobenius norm.} \be{Poonformula} \left\| A^n -B^n
\right\| \leq n \left\| A - B \right\|, \ee we write \be{forerr}
\tilde Err= < 0.235 +10^5 \left\|    T_5(x)- \tilde T_5(x,x_1)
\right\|.  \ee In view of our bound on the error of $0.4$, we need
to find $x_1>0$ so that $\left\|    T_5(x)- \tilde T_5(x,x_1)
\right\| \leq 0.165 \times 10^{-5}$. Now, we have \be{normeinequa}
\left\| T_5 - \tilde T_5 \right\|= \left\|e^{-A_1 x^{\frac{1}{4}}}
\Pi_1 e^{-A_1 x^{\frac{1}{4}}} \Pi_2 - e^{A_1 x_1} \Pi_1 e^{A_1 x_1}
\Pi_2 \right\| = \left\| \Pi_1 - e^{A_1 (x^{\frac{1}{4}}+x_1)} \Pi_1
e^{A_1 (x^{\frac{1}{4}}+x_1)} \right\|.
 \ee
Therefore we have \be{contnormineq} \left\| T_5 - \tilde T_5
\right\| \leq \left\|  \Pi_1 - e^{A_1 (x^{\frac{1}{4}}+x_1)} \Pi_1
\right\| \ee $$+ \left\| e^{A_1 (x^{\frac{1}{4}}+x_1)}  \Pi_1 -
e^{A_1 (x^{\frac{1}{4}}+x_1)} \Pi_1 e^{A_1 (x^{\frac{1}{4}}+x_1)}
\right\| \ =2 \left\| {\bf 1}- e^{A_1 (x^{\frac{1}{4}}+x_1)}
\right\|.
$$
Therefore, we need to find $x_1 \geq 0$ so that \be{pprr} \| {\bf 1}
- e^{A_1 (x^{\frac{1}{4}}+x_1)} \| \leq \frac{0.165 \times
10^{-5}}{2}.   \ee We calculate explicitly the eigenvalues of $A_1$
which are given by $\pm i r$ and $\pm il$, with $r$ and $l$ given by
\be{erreeelle} r:=\sqrt{\frac{15+ \sqrt{125}}{2}}, \qquad l:=
\sqrt{\frac{15 - \sqrt{125}}{2}}.   \ee We  have \be{klop} \| {\bf
1} - e^{A_1 (x^{\frac{1}{4}}+x_1)} \|=2\sqrt{1-
\cos(r(x^{\frac{1}{4}}+x_1))+  1- \cos(l(x^{\frac{1}{4}}+x_1))}. \ee
Therefore,  setting $t:=x^{\frac{1}{4}}+x_1$, formula (\ref{pprr})
is certainly satisfied if \be{condi1} 1-\cos(rt) < 8 \times
10^{-14},  \ee and \be{condi2} 1-\cos(lt) < 8 \times 10^{-14}.  \ee
Setting $\epsilon:=\arccos(1-8 \times 10^{-14})$, we need to find $t
> x^{\frac{1}{4}}$, positive integers $p$ and $q$ such that
\be{fth} \left| rt -2\pi p \right| <\epsilon, \qquad   \left| lt
-2\pi q \right| <\epsilon. \ee One way to do this is as follows. Fix
an integer $k >0$ large enough so that \be{ght} \frac{1}{k} <
\frac{\epsilon}{2 \pi}.  \ee According to Dirichlet's approximation
theorem (see, e.g., \cite{RNG} Theorem 1.3) we can find $p$ and $q$,
with $1 \leq p \leq k$ positive integers so that \be{pkpkpk} \left|
\frac{l}{r} p -q \right| < \frac{1}{k}. \ee Choose $p$ and $q$ this
way and \be{t} t=\frac{2 \pi p}{r}.  \ee Using this value of $t$,
the first one of (\ref{fth}) is verified because the left hand side
becomes zero. Replacing this value of $t$ in the second one of
(\ref{fth}) and using (\ref{ght}) and (\ref{pkpkpk}) we obtain that
the second inequality is satisfied as well. Moreover, since $q \geq
1$, we have that \be{kla} t \geq \frac{2 \pi}{r}\approx 1.7366
> x^{\frac{1}{4}}= \left( \frac{\pi}{44} 10^{-5}
\right)^{\frac{1}{4}} \approx 0.0291.  \ee This concludes the
example.

\section{Combination of the two methods}
\label{M3} The main ideas in the two methods of control described in
the previous sections can be combined in a third method. The main
idea of the method in Section \ref{M1} was to use similarity
transformation to generate a basis of the dynamical Lie algebra
${\cal L}$ starting from the given matrices in ${\cal F}$ in
(\ref{effematrices}) (cf. (\ref{effe})). The main idea of the method
in section \ref{M2} is the use of the limit in Lemma
\ref{Limitefondamentale}, once (\ref{AIG}) holds. This   allows us
to control to the target, by repeating a given sequence of available
exponentials, with arbitrary accuracy. We can combine the two ideas.
We  first use similarity transformations to obtain a basis of ${\cal
L}$, ${A}_1,\ldots,A_{\dim{\cal L}}$. Then, if $e^{H}$ is the target
and $H=\sum_{j=1}^{\dim{\cal L}} \alpha_j A_j$, we  use the fact
that \be{xfg} e^{Hx}=e^{\sum_{j=1}^{\dim{\cal L}} \alpha_j
A_j}=\prod_{j=1}^{\dim{\cal L}} e^{\alpha_j A_j x}+O(x^2),  \ee
along with Lemma \ref{Limitefondamentale} to approximate with
arbitrary accuracy the target state, i.e., \be{furthrd}
e^{H}=\lim_{n \rightarrow \infty} \left[ \prod_{j=1}^{\dim{\cal L}}
e^{\alpha_j A_j \frac{1}{n}} \right]^n.\ee At the end of the
process, we replace all the exponentials of the form $e^{A_j t}$
with $t < 0$ with approximating exponentials of the form $e^{A_j
\bar t}$ with $\bar t >0$.

\vs

We test this method on  the example in subsection \ref{subsec}.
Given $A_1$ and $A_2$ in (\ref{calfexample}) we calculate
\be{effeex} F:=e^{A_2 \frac{\pi}{2}} A_1
e^{-A_2\frac{\pi}{2}}=\pmatrix{0 & -1 & 0 & 2 \cr 1 & 0 & 1 & 0 \cr
0 & -1 & 0 & -3 \cr -2 & 0 & 3 & 0}.  \ee Our target is $e^{A_5
\frac{\pi}{44}}$ in (\ref{pog}). We have the decomposition
\be{decoA5} A_5=10 A_1+6F-16 A_2,  \ee so that \be{gfa} e^{A_5
\frac{\pi}{44}x}=R_5(x)+O(x^2),  \ee with \be{R5xxxx} R_5(x):= e^{10
A_1 \frac{\pi}{44}x} e^{6 F \frac{\pi}{44}x} e^{-16 A_2
\frac{\pi}{44}x}.  \ee We have, according to Lemma
\ref{Limitefondamentale}, \be{LMF} \lim_{n \rightarrow \infty}
 \left[ R_5\left(\frac{1}{n}\right) \right]^n=e^{A_5 \frac{\pi}{44}}. \ee Table
\ref{tavola2} shows the results of numerical experiments with this
scheme displaying  the error $Err$ as a function of the number of
iterations. Compared with Table \ref{tavola}, it is clear that this
method converges much faster. Another advantage is that the all the
exponentials $e^{At}$ with negative $t$ are for  $A=A_2$ (cf.
(\ref{R5xxxx}) and
  (\ref{effeex})) and the one dimensional
subgroup associated  with $A_2$ is closed. Therefore no further
approximation is needed.

\begin{table}[ht]
\caption{Results of numerical experiments for the method in section
\ref{M3}.} \vs \centering
\begin{tabular}{c c}
\hline \hline number of iterations $n$ & Error $Err$\\
\hline 2 & 2.2819 \\
\hline 10 & 0.4544 \\
\hline 20 & 0.2267 \\
\hline  50 & 0.0906 \\
\hline 100 & 0.0453 \\
\hline 1000 & 0.0045 \\
\hline 10000 & 0.0005 \\
\label{tavola2}
\end{tabular}
\end{table}

\section{Conclusions}

The methods described in this paper can be seen as a constructive
proof of the Lie algebra rank condition of Theorem \ref{LARC}.  It
is expected that the ideas described above  can be extended and
improved by using more sophisticated exponential formulas
\cite{Y25}, in many ways. It is also expected that it will be
possible to obtain estimates of the convergence rate in various
cases. Our goal here was to propose ideas that, although at an early
stage, are very general and, in principle,  allow us  to control
every system on a compact Lie group. These systems in particular
include  the important class of closed, finite dimensional, quantum
systems which are coherently controlled, namely controlled through a
change in the Hamiltonian.

\vs

In the future, it will be important to improve the algorithms by
minimizing the number of switches in the control laws that mainly
depends on the number of iterations, in the last two sections. In
this respect, the algorithm of section \ref{M3} is expected to be
faster than the algorithm in section \ref{M2}, as a consequence of
the exponent $2$, in the $O(x^2)$ in (\ref{xfg}) as opposed to the
exponent $1+\delta$ (with $\delta$ typically $ < 1$) in (\ref{AIG}).
If our main concern is however the time of implementation,  the
effect of an increasing number of iterations $n$ in (\ref{polp}) is
balanced by the $\frac{1}{n}$ exponents inside the limit. The main
problem, in terms of time,  is the approximation of matrices of the
form $e^{At}$ with $t < 0$ with matrices of the form $e^{At}$, with
$t >0$,  in the case of non-closed subgroups. In fact, we might have
to `travel' for a long time inside the Lie group $e^{\cal L}$ before
we get close enough to the original $e^{At}$. In special situations,
however, it might be  possible to transform $A$ into $-A$ via
available similarity transformations, or reduce ourselves to a
smaller dimensional Lie subgroup where the problem is more easily
tractable. Nevertheless, it is always possible to find such an
approximation and therefore the control. Remark \ref{nt} shows how
this problem can be reduced to a standard problem of Diophantine
approximation in number theory for which there exist a vast
literature and that can be always solved in principle.

In conclusion,  would like to comment on the assumption of
compactness which is used in the paper only in two instances. In
particular, compactness is used to have a surjective exponential map
and to be able to approximate an exponential of the form $e^{At}$
with $t$ negative with an exponential of the same type with $t$
positive. Whenever these two properties hold, the methods of this
paper can still be applied to more general Lie groups. In particular
this is the case for finite dimensional closed quantum mechanical
systems whose dynamical Lie algebra $\cal L$ is a subalgebra of
$u(n)$.

\vs

\vs

\vs

{\bf Acknowledgment} This research was supported by NSF under Grant
ECCS0824085. The author would like to thank Y.T. Poon for suggesting
formula (\ref{Poonformula}). He also would like to thank Richard Ng
for indicating the relevant literature on Diophantine approximation
and Dirichlet's approximation theorem, and for helpful discussions
on this topic.

\end{document}